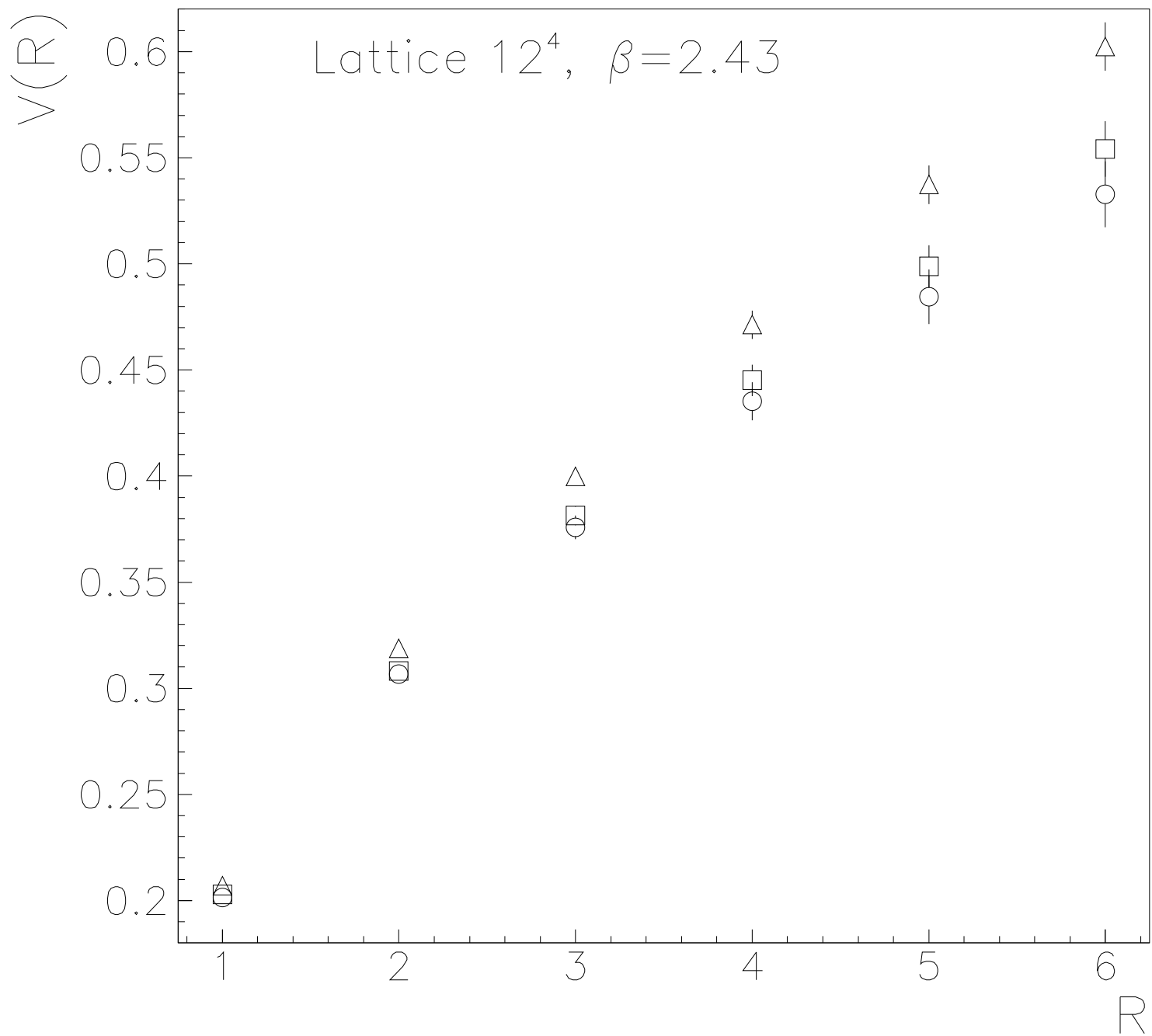

# New algorithm for gauge fixing in SU(2) lattice gauge theory *

G.S. Bali[a][†], V. Bornyakov[b][‡], M. Müller-Preussker[c] and F. Pahl[d]

[a]Fachbereich Physik, Bergische Universität, Gesamthochschule, 42097 Wuppertal, Germany

[b]Institute of High Energy Physics, 142284 Protvino, Russia

[c]Institut für Physik, Humboldt Universität zu Berlin, 10099 Berlin, Germany

[d]Fachbereich Physik, Freie Universität, 14195 Berlin, Germany

An overrelaxed variant of simulated annealing is applied to the problem of maximally abelian gauge fixing. The superiority of this algorithm over the commonly used relaxation procedure is demonstrated. Biases on non gauge invariant quantities due to gauge fixing ambiguities are discussed.

## 1. INTRODUCTION

Though gauge fixing can generally be avoided in LGT, many physically interesting non-local quantities explicitly depend on the gauge. Prominent examples are gluon and quark propagators. Here, we address investigations of the dual superconductor scenario of confinement (for details see e.g. Ref. [1]). Maximally abelian (MA) gauge is believed to be most suitable for such studies. Fixing the MA gauge amounts to maximizing the functional ($V = N_{\text{sites}}$)

$$F(U) = \frac{1}{8V} \sum_{n,\mu} \text{tr}\left(\sigma_3 U_{n,\mu} \sigma_3 U_{n,\mu}^\dagger\right) \quad , \qquad (1)$$

with respect to local gauge transformations,

$$U_{n,\mu} \to U_{n,\mu}^g = g_n U_{n,\mu} g_{n+\mu}^\dagger, \; g_n \in SU(2)/U(1) \quad (2)$$

Besides absolute maxima – which can be degenerate even beyond global gauge transformations and local ones within the remaining ungauged $U(1)$ subgroup – the function $F(U)$ can have numerous local maxima. This feature resembles the Gribov problem in continuum gauge theories [2]. Whereas degenerate global maxima, at least for Landau gauge, turn out to be physically equivalent [3], local maxima may lead to wrong physical results [4]. Therefore, in numerical practice one would like to find local maxima as close as possible to the absolute ones. In this way we hope for smaller systematic uncertainties due to gauge fixing ambiguities.

So far the relaxation plus overrelaxation (RO) algorithm has been employed for MA gauge fixing. In this paper we report on an implementation of an algorithm which is successfully used in various optimization problems, the simulated annealing (SA) algorithm [5,6]: the functional $F(U)$ is regarded as a "spin action",

$$S(s) = F(U^g) = \frac{1}{8V} \sum_{n,\mu} \text{tr}\left(s_n U_{n,\mu} s_{n+\mu} U_{n,\mu}^\dagger\right) \quad (3)$$

where $s_n = g_n^\dagger \sigma_3 g_n$ resemble the spin variables. The lattice fields $U_{n,\mu}$ play the role of (almost) random local couplings. Maximizing the functional $F(U^g)$ is equivalent to decreasing the auxiliary temperature $T$ of the statistical system with partition function

$$Z = \sum_{\{s_n\}} \exp\left(\frac{1}{T} S(s)\right) \quad . \qquad (4)$$

One starts with equilibrating this spin glass at high temperature. Subsequently, $T$ is gradually decreased keeping the system very close to equilibrium. It is evident that in the limit $T \to 0$ the system approaches its ground state, i.e. the maximal value of $S$. The advantage of SA can be formulated in terms of solid state physics: standard relaxation corresponds to fast cooling and causes defects while the adiabatic cooling procedure of SA avoids these defects. In order to improve the movement of the spin variables through phase space we combine simulated annealing with overrelaxation (OSA).

---

*TALK GIVEN AT THE LATTICE '94 INTERNATIONAL SYMPOSIUM ON LATTICE FIELD THEORY, BIELEFELD, GERMANY, SEPTEMBER 27 – OCTOBER 1, 1994
[†]Supported by EU project SC1*-CT91-0642.
[‡]Supported by DFG grant 436-RUS-113 during completion of this work.

## 2. SA IMPLEMENTATION

Our procedure consists of three steps: 1) thermalization at $T = 2.5$; 2) gradual decrease of $T$ down to $T = 0.01$; 3) final maximization by means of the RO algorithm.

In steps 1 and 2 an overrelaxation transformation is performed at six consecutive lattice sites and heatbath is applied to the seventh. Within step 2, every time when the heatbath update is applied to a site, the temperature is decreased by a quantum $\delta T$. A version of this algorithm where sites have been visited in lexicographical ordering, within subcubes of $2^4$ sites each, in parallel has also been tested successfully. The combination of local overrelaxation and local temperature decreasing enables us to reduce the absolute number of sweeps required within the cooling process while remaining close to an equilibrium state.

Within the temperature range $2.5 \geq T \geq 0.1$, $\delta T(T)$ has been tuned such that the spin action increases about linearly with the number of iteration sweeps. This has been realized by subdividing this range into 24 intervals of width $\Delta T = 0.1$. The corresponding differences of the action $\Delta S(T) = S(T) - S(T - \Delta T)$ have been computed on equilibrated configurations. $\Delta S(T)$ was found to be very stable against statistical fluctuations among different Monte Carlo configurations. An almost undetectable volume dependence and a moderate dependence on $\beta$ was noticed. The number of sweeps (out of a fixed total number) to be performed within each interval $[T - \Delta T, T]$ was computed to be proportional to $\Delta S(T)$ and, subsequently, the corresponding value of $\delta T(T)$ has been determined. Within the region $0.1 \geq T \geq 0.01$, 50 additional sweeps have been performed. Finally, the RO algorithm has been applied till a convergence criterion was satisfied (about 130 sweeps in average).

## 3. RESULTS AND DISCUSSION

Our main simulations have been performed on a $12^4$ lattice at $\beta = 2.43$. We studied the SA algorithm extensively, using three different cooling schedules at step 2 of our procedure: 250 (OSA1), 500 (OSA2) and 1000 (OSA3) total sweeps. In addition, the standard procedure (RO) has been applied with the same convergence criterion. We collected 30 statistically independent equilibrated MC configurations. On each of these configurations, 10 random gauge copies have been generated. Each of the four algorithms has been applied to these copies.

Let $A$ be a gauge dependent abelian quantity. In the following we abbreviate $\overline{A}$ as the average over gauge copies and $\langle A \rangle$ as the statistical average. In table 1, results of the four algorithms for various quantities are compared with each other. $\sigma_{sd}^2 = \langle \overline{F^2} - \overline{F}^2 \rangle$ denotes the variation of the maximized value of the functional among gauge copies. The ideal algorithm would always give $F = F_{\max}$, i.e. $\sigma_{sd} = 0$. $\rho_m$ is the monopole density. $W_{ii}$ are abelian $i \times i$ Wilson loops and $K^{ab}$ is the abelian string tension. A comparison of $\langle \overline{F} \rangle$ and $\sigma_{sd}$ exhibits the superiority of all OSA schedules over the standard RO algorithm, the longest schedule yielding the best results. Differences between the OSA2 and OSA3 algorithms are on the level of statistical errors. In means of total computer time spent, OSA2 is about a factor two slower than RO. OSA1, still being an improvement, is only 30 % slower than RO. (On a CM-5, OSA1 was found to be even slightly faster than RO.)

It is obvious from the table that physical results dramatically depend on the care spent on gauge fixing. The functional is correlated with Wilson loops and anti-correlated with the monopole number and abelian string tension.

By applying state-of-the-art smearing techniques on the spatial transporters of abelian Wilson loops, we have been able to compute the abelian ground state potential. In fig. 1 these po-tentials are displayed for the RO (used by all previous authors for MA gauge fixing) and the OSA3 algorithms. In addition the copy with largest functional among all 40 copies has been chosen on each configuration as our best estimate of the "true" maximum (circles). From this potential we obtain $K^{ab}_{best} = .0478(38)$ as an estimate of the abelian string tension. (All errors have been obtained by the jackknife procedure.) From fig. 1 it is evident that this value will be overestimated by about 30% by use of the standard procedure (see also table 1). Even our most expensive algorithm OSA3 yields a value that is off by about one statistical standard deviation ($K^{ab}_{OSA3} = .0536(30)$).

The question of practical importance is how to obtain a string tension estimate which is reasonably close to $K^{ab}_{best}$ at lowest cost. In order to answer this question the string tension has been evaluated for each algorithm using the copy with largest value of $F$ out of 2, 3, etc. gauge copies. We came to the following conclusion: to reproduce $K^{ab}_{best}$ with a systematic error of less than 5% one can either use the OSA3 schedule with 2 copies per configuration or the OSA2 schedule with 3 gauge copies which is slightly faster in terms of computer time.

In order to study the scaling properties, the investigations have been partly repeated on a lattice with about the same physical size but smaller lattice resolution ($16^4$ at $\beta = 2.51$). The OSA2 and RO algorithms have been applied to 10 copies on 20 configurations. The qualitative properties are the same. Again, the string tension obtained after application of the RO algorithm is dramatically overestimated: $K^{ab}_{RO} = .0362(15)$ versus $K^{ab}_{best} = .0305(19)$. Since the non abelian string tension at this $\beta$ value comes out to be [7] $K = .0337(5)$, this difference is relevant to the physical conclusions drawn.

## 4. CONCLUSIONS

It has been demonstrated that the quality of MA gauge fixing (in terms of the value of $F(U^g)$ and the scatter of results among different gauge copies) can be improved by applying the OSA algorithm even without any draw back in computer time. Our systematic study of non gauge invariant quantities revealed the importance of such an improvement and should be understood as a warning, not to extract the abelian potential and other observables without carefully investigating systematic errors, induced by the gauge fixing procedure. The nature of the ambiguities deserves further study. An extension of the method to other gauges (e.g. Landau gauge) is straight forward.

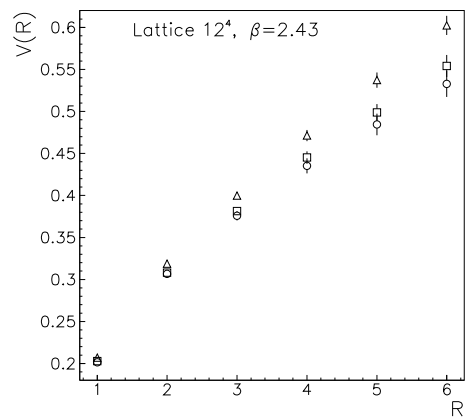

Figure 1. Abelian potentials from RO (triangles), OSA3 (squares) and "best" copy (circles).

Table 1. Comparison of gauge fixing algorithms.

|  | RO | OSA1 | OSA2 | OSA3 |
| --- | --- | --- | --- | --- |
| $\langle F \rangle$ | 0.7370(2) | 0.7383(2) | 0.7387(2) | 0.7390(2) |
| $\sigma_{sd}$ | $18 \cdot 10^{-5}$ | $13 \cdot 10^{-5}$ | $10 \cdot 10^{-5}$ | $8 \cdot 10^{-5}$ |
| $\rho_m$ | 0.0218(2) | 0.0209(3) | 0.0207(3) | 0.0204(3) |
| $W_{11}$ | 0.7702(5) | 0.7716(5) | 0.7720(5) | 0.7723(5) |
| $W_{44}$ | 0.085(1) | 0.090(1) | 0.091(1) | 0.092(2) |
| $K^{ab}$ | 0.063(3) | 0.057(3) | 0.055(3) | 0.054(3) |